\PassOptionsToPackage{style=ieee, citestyle=numeric-comp, maxnames=99, sortcites=true, backend=bibtex}{biblatex}
\PassOptionsToPackage{scaled=0.8}{inconsolata}

\documentclass[sigconf,natbib=false]{acmart}\def\UseACMTemplate{1}

\usepackage{graphicx}
\usepackage{xcolor}
\definecolor{B}    {HTML}{000000}   
\definecolor{B2}   {HTML}{003399}   
\definecolor{Bv}   {HTML}{0000EB}   
\definecolor{R}    {HTML}{c9171e}
\definecolor{R2}   {HTML}{d7003a}
\definecolor{INK}  {HTML}{595857}
\definecolor{Y}    {HTML}{f1c40f}
\definecolor{G}    {HTML}{009a00}
\definecolor{GRAY} {HTML}{808080}
\definecolor{MAUVE}{HTML}{9400D1}
\definecolor{crimson}{rgb}{0.86, 0.08, 0.24}
\usepackage{amsmath,amsfonts}    
\usepackage{mathtools}
\usepackage{bm}
\usepackage{amsthm}

\usepackage{array} 
\usepackage{booktabs}
\usepackage{colortbl}
\usepackage{adjustbox}
\usepackage{multirow}

\usepackage{url, hyperref}                  
\hypersetup{
    pdfborder={0 0 0}}

\usepackage{enumitem}                       
\usepackage{lipsum}                         
\usepackage[plain]{algorithm}       		
\floatstyle{plaintop}
\usepackage{algpseudocode}
\algrenewcommand{\alglinenumber}[1]{{\scriptsize\bfseries\ttfamily\color{R}#1}}
\floatstyle{plaintop}
\restylefloat{algorithm}

\usepackage{xpatch}
\makeatletter
\xpatchcmd{\algorithmic}{\ALG@tlm\z@}{\ALG@tlm\z@\leftmargin 10pt}{}{}
\makeatother

\usepackage{listings}						
\usepackage{caption, subcaption}            
\usepackage{setspace}                       

\usepackage{soul}
\ifx\UseACMTemplate\undefined
\else
    \newcommand{\SEC}{\textcolor{black}{\S}}
    \newcommand{\FIG}{\textcolor{black}{Figure}}
    
\fi

\ifx\UseIEEETemplate\undefined
\else
    \newcommand{\SEC}{\textcolor{black}{\S}}
    \newcommand{\FIG}{\textcolor{black}{Fig.}}
    
\fi

\usepackage{comment}

\newcommand{\BOLD}{\fontfamily{ugq}\selectfont}

\usepackage{tikz}
\usetikzlibrary{positioning}        
\usetikzlibrary{shapes.multipart}   
\usetikzlibrary{shapes.geometric}   
\usetikzlibrary{arrows}
\usetikzlibrary{arrows.meta}        
\usetikzlibrary{backgrounds}
\usetikzlibrary{patterns}
\usetikzlibrary{fit}
\usetikzlibrary{shadows.blur}
\usetikzlibrary{shapes.symbols}

\usepackage{wrapfig}

\newcommand*\Circled[1]{
	\tikz[baseline=(char.base)]{%
		\node[shape=circle, draw=none, fill=gray!40, thick, inner sep=0.6pt] (char) {%
			\textcolor{black}{\sffamily#1}}; }}

\newcommand{\TABLETITLE}{\BOLD\color{black}}
\newcommand{\TABLECAPTION}{\color{R}\scriptsize}

\usepackage{biblatex}
\newcommand{\cusz}{\textsc{cuSZ}}       
\newcommand{\cuzfp}{cuZFP}       

\newcommand{\cudaSZx}{cuSZx}
\newcommand{\thiswork}{FZ-GPU}
\newcommand{\cudamgard}{MGARD-GPU}
\renewcommand{\cusz}{\textsc{cuSZ}}

\copyrightyear{2023} 
\acmYear{2023} 
\setcopyright{usgovmixed}\acmConference[HPDC '23]{Proceedings of the 32nd International Symposium on High-Performance Parallel and Distributed Computing}{June 16--23, 2023}{Orlando, FL, USA}
\acmBooktitle{Proceedings of the 32nd International Symposium on High-Performance Parallel and Distributed Computing (HPDC '23), June 16--23, 2023, Orlando, FL, USA}
\acmPrice{15.00}
\acmDOI{10.1145/3588195.3592994}
\acmISBN{979-8-4007-0155-9/23/06}

\begin{document}
\title[\thiswork]{\thiswork: A Fast and High-Ratio Lossy Compressor for Scientific Computing Applications on GPUs}

\settopmatter{authorsperrow=4}

\newcommand{\wsu}{Washington State University}
\newcommand{\iu}{Indiana University}
\newcommand{\ua}{The University of Alabama}
\newcommand{\ucr}{University of California, Riverside}
\newcommand{\clemson}{Clemson University}
\newcommand{\anl}{Argonne National Laboratory}
\newcommand{\ornl}{Oak Ridge National Laboratory}
\newcommand{\unt}{University of North Texas}
\newcommand{\uky}{University of Kentucky}

\newcommand{\AFFIL}[4]{%
    \affiliation{%
        \institution{\small #1}
        \city{#2}\state{#3}\country{#4}
    }
}

\newcommand{\IU}{\AFFIL{\iu}{Bloomington}{IN}{USA}}
\newcommand{\WSU}{\AFFIL{\wsu}{Pullman}{WA}{USA}}
\newcommand{\UA}{\AFFIL{\ua}{Tuscaloosa}{AL}{USA}}
\newcommand{\UCR}{\AFFIL{\ucr}{Riverside}{CA}{USA}}
\newcommand{\ANL}{\AFFIL{\anl}{Lemont}{IL}{USA}}
\newcommand{\ORNL}{\AFFIL{\ornl}{Oak Ridge}{TN}{USA}}
\newcommand{\CLEMSON}{\AFFIL{\clemson}{Clemson}{SC}{USA}}
\newcommand{\UKY}{\AFFIL{\uky}{Lexington}{KY}{USA}}
\newcommand{\UNT}{\AFFIL{\unt}{Denton}{TX}{USA}}

\author{Boyuan Zhang}{\IU}
\email{bozhan@iu.edu}
\authornote{Boyuan Zhang and Jiannan Tian are co-first authors.}

\author{Jiannan Tian}{\IU}
\email{jti1@iu.edu}
\authornotemark[1]

\author{Sheng Di}{\ANL}
\email{sdi1@anl.gov}

\author{Xiaodong Yu}{\ANL}
\email{xyu@anl.gov}

\author{Yunhe Feng}{\UNT}
\email{yunhe.feng@unt.edu}

\author{Xin Liang}{\UKY}
\email{xliang@uky.edu}

\author{Dingwen Tao}{\IU}
\authornote{Corresponding author: Dingwen Tao, Department of Intelligent Systems Engineering, Luddy School of Informatics, Computing, and Engineering, Indiana University.}
\email{ditao@iu.edu}

\author{Franck Cappello}{\ANL}
\email{cappello@mcs.anl.gov}

\renewcommand{\shortauthors}{Zhang \& Tian et~al.}

\begin{abstract}
	Today's large-scale scientific applications running on high-perfor\-mance computing (HPC) systems generate vast data volumes. Thus, data compression is becoming a critical technique to mitigate the storage burden and data-movement cost. However, existing lossy compressors for scientific data cannot achieve a high compression ratio and throughput simultaneously, hindering their adoption in many applications requiring fast compression, such as in-memory compression. To this end, in this work, we develop a fast and high-ratio error-bounded lossy compressor on GPUs for scientific data (called \thiswork). Specifically, we first design a new compression pipeline that consists of fully parallelized quantization, bitshuffle, and our newly designed fast encoding. Then, we propose a series of deep architectural optimizations for each kernel in the pipeline to take full advantage of CUDA architectures. We propose a warp-level optimization to avoid data conflicts for bit-wise operations in bitshuffle, maximize shared memory utilization, and eliminate unnecessary data movements by fusing different compression kernels. Finally, we evaluate \thiswork{} on two NVIDIA GPUs (i.e., A100 and RTX A4000) using six representative scientific datasets from SDRBench. Results on the A100 GPU show that \thiswork{} achieves an average speedup of 4.2$\times$ over \cusz{} and an average speedup of 37.0$\times$ over a multi-threaded CPU implementation of our algorithm under the same error bound. \thiswork{} also achieves an average speedup of 2.3$\times$ and an average compression ratio improvement of 2.0$\times$ over \cuzfp{} under the same data distortion.
\end{abstract}



\maketitle

\section{Introduction}
\label{sec:intro}

\paragraph*{Motivation} Large-scale scientific applications running on high-performance computing (HPC) systems produce vast data for post hoc analysis. For instance, Hardware/Hybrid Accelerated Cosmology Code (HACC)~\cite{hacc,miraio} may produce petabytes of data in hundreds of snapshots when simulating one trillion particles.
Storing such a large amount of data could be vastly inefficient, especially to parallel file systems (PFS) with relatively low I/O bandwidth ~\cite{liang2018error,xincluster18}.

Data reduction is becoming an effective method to resolve the big data issue in scientific applications.
Although traditional lossless data reduction methods such as data deduplication and lossless compression can guarantee no information loss, they suffer from limited compression ratios on scientific datasets.
Specifically, deduplication usually reduces the scientific data size by only 20\% to 30\%~\cite{meister2012study}, and lossless compression achieves a compression ratio of up to $\sim$2:1~\cite{son2014data}.
However, the data reduction ratios provided by these methods are much lower than the ratios scientists desire~\cite{use-case-Franck}.

Error-bounded lossy compressors have been studied for years to address this issue for scientific data reduction.
Not only can they achieve very high compression ratios (e.g., over 100$\times$)~\cite{sz16,sz17,zfp,liang2018error}, but they can also strictly control data distortion concerning user-set error bounds.
Notably, a satisfying lossy compressor designed for scientific data reduction should address three primary concerns simultaneously: \Circled{1} high compression ratio, \Circled{2} high throughput, and \Circled{3} high compression quality (data fidelity).
Most of the existing error-bounded lossy compressors (such as SZ~\cite{sz16,sz17,sz3}, FPZIP~\cite{fpzip}, ZFP~\cite{zfp}), however, are mainly designed for CPU architectures, which cannot meet the high-throughput requirement.
For example, X-ray imaging data generated on advanced instruments such as LCLS-II laser~\cite{lcls} can result in a data acquisition rate of 250~GB/s~\cite{use-case-Franck}.
As such, high compression throughput is essential to store large amounts of data for scientific projects efficiently.

\vspace{-1mm}
\paragraph*{Limitations of state-of-the-art approaches.}
Existing error-bounded lossy compressors for GPUs (such as {\cusz}~\cite{cusz2020}, cuZFP
\cite{cuZFP}, and \cudamgard{}~\cite{chen2021accelerating}) suffer from either low throughputs or low compression ratios.
Specifically, although cuZFP has slightly higher throughput compared with {\cusz} and \cudamgard, it supports only the fixed-rate mode~\cite{compression-mode}, which suffers much lower compression quality than the fixed-accuracy mode (a.k.a error-bounded mode)~\cite{jin2020understanding}, significantly limiting its adoption in practice.
On the other hand, {\cusz} and \cudamgard{} can achieve much higher compression ratios than cuZFP, but their compression throughputs are relatively low.
This is because both MGARD and SZ algorithms require entropy and dictionary encoding to achieve high compression ratios due to the aggregate repeated symbols (e.g., quantization codes generated by the prediction and quantization stages in SZ). At the same time, \Circled{1} \cudamgard{} uses DEFLATE~\cite{deutsch1996rfc1951} (including Huffman entropy encoding~\cite{Huffman-original} and LZ77 dictionary encoding~\cite{ziv1977universal}) on the CPU, causing low throughput, and \Circled{2} {\cusz} adopts an inefficient GPU-based Huffman encoding~\cite{cusz2020}.

Specifically, Huffman encoding and most dictionary encoding algorithms contain substantial data dependencies, making them difficult to parallelize on GPUs extensively.
Moreover, designing an efficient parallel dictionary encoding is challenging because of the intrinsic dependency in its repeated sequence search.
Thus, {\cusz} leaves this part to the CPU, which incurs high time overhead, including data transfer.
Furthermore, achieving high performance on GPUs requires maximizing the parallelism of GPU threads and using shared memory and mitigating issues of coherence, warp divergence, and bank conflicts.
Thus, it is challenging to develop an efficient GPU-based error-bounded lossy compressor that simultaneously achieves a high compression ratio and high throughput.

\vspace{-1mm}
\paragraph*{Key insights and contributions.} In this work, we propose a fast and high-ratio error-bounded lossy compressor (called \thiswork~\footnote{The code is available at \url{https://github.com/szcompressor/FZ-GPU}.}) for scientific computing applications on GPU based on the {\cusz} framework~\cite{cusz2020}, which maximizes the overall throughput.
Specifically, we first propose to use bitshuffle~\cite{masui2017bitshuffle} to rearrange the quantization codes generated by the prediction-and-quantization step (called ``dual-quantization'') in {\cusz} (will be introduced in \SEC\ref{sec:cusz}) at the bit level to increase the data correlation for more effective encoding.
We then design a new fast GPU lossless encoding method for bitshuffled data.
We carefully design a GPU kernel to fuse bitshuffle and encoding operations to reduce unnecessary data movements between global and shared memories.
Using the proposed bitshuffle and encoding approach, we can eliminate the inefficient Huffman encoding from {\cusz}.
Moreover, we optimize the performance of the dual-quantization method by eliminating the outliers handling mechanism and the data shift operation, hence improving the effectiveness of the subsequent bitshuffle process.

The main contributions of our work are summarized as follows.
\begin{itemize}[noitemsep, topsep=6pt, leftmargin=1.3em]
	\item We propose a new compression pipeline based on the {\cusz} framework, which consists of our optimized dual-quantization, bitshuffle, and our proposed lossless encoding after bitshuffle (to replace the slow Huffman encoding implementation) on GPUs.
	\item We optimize the dual-quantization by eliminating the data-shift and outlier-handling operations, improving both compression throughput and bitshuffle efficiency, thus increasing the compression ratio.
	\item We develop a GPU bitshuffle kernel with a warp-level optimization to avoid data conflicts in bit-wise operations. We also maximize the utilization of shared memory to improve the performance of this memory-intensive kernel.
	\item We design a new lossless encoding method that leverages the data characteristics after bitshuffle and the high parallelism of GPUs. It can effectively and efficiently remove the high redundancy from bitshuffle, thereby achieving a high compression ratio and throughput.
	\item We carefully fuse the bitshuffle kernel and the first phase of the encode kernel (i.e., recording zero blocks) into a single GPU kernel to eliminate unnecessary data movements between the GPU's global and shared memory.
	\item We evaluate \thiswork{} on six real-world scientific application datasets from \emph{Scientific Data Reduction Benchmarks}~\cite{zhao2020sdrbench} on two NVIDIA GPUs (i.e., A100 and RTX A4000) and compare it to four state-of-the-art compressors. Experiments show that on the A100 GPU, \thiswork{} significantly improves the compression throughput by up to 11.2$\times$ over {\cusz}; and compared to cuZFP, \thiswork{} achieves an average of 2.0$\times$ higher compression ratio at the same data distortion with an average speedup of 2.3$\times$.
\end{itemize}

\vspace{-1mm}
\paragraph*{Limitations of the proposed approach.}
Compared to {\cusz}, \thiswork{} significantly improves the compression throughput in all cases, while it has a slightly lower compression ratio at low error bounds.
Compared to \cudaSZx, \thiswork{} has much higher compression ratios and hence higher overall data-transfer throughput, but its compression throughput is lower than \cudaSZx.
Compared to \cuzfp, \thiswork{} has a slightly lower compression ratio and throughput at large error bounds (e.g., over 5e$-$3) in some datasets.

The remaining of this paper is organized as follows.
In \SEC\ref{sec:sz-bg}, we present the background about GPU lossy compression for scientific data, \cusz{} framework, prediction-and-quantization method, and our problem statement. In \SEC\ref{sec:design}, we describe the design of our proposed \thiswork. In \SEC\ref{sec:eval}, we evaluate \thiswork{} on different scientific datasets and compare it with other compressors. In \SEC\ref{sec:related}, we discuss related work on GPU-based lossy compression. Finally, in \SEC\ref{sec:conclusion}, we conclude our work and discuss future work.

\section{Background and Problem Statement}
\label{sec:sz-bg}

In this section, we introduce the background about GPU-based lossy compression and cuSZ framework and our problem statement.

\begin{figure*}[ht]

	\sffamily
	\centering

	\includegraphics[width=\linewidth]{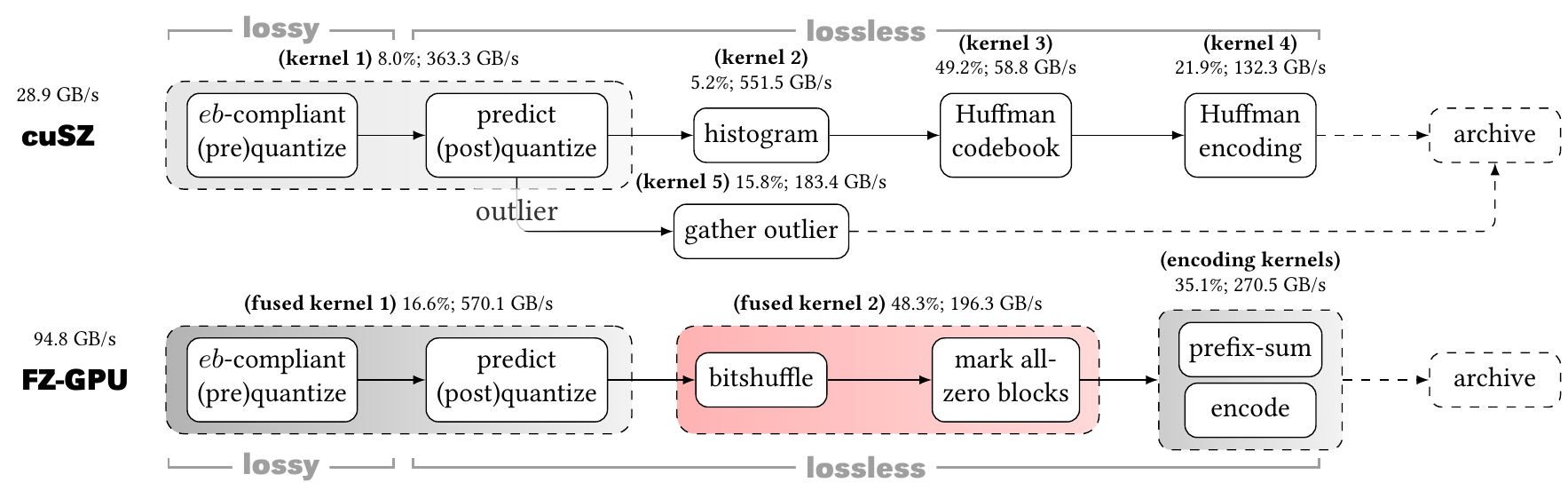}

	\caption{Our proposed new compression pipeline (``\thiswork'') versus original \cusz's compression pipeline.
			{Each kernel is marked with its relative time (in percentage) and throughput (in GB/s) based on one field from the Hurricane dataset at an error bound of 1e-4.}}
	\vspace{-4mm}
	\label{fig:system}
\end{figure*}

\subsection{GPU Lossy Compression for Scientific Data}

There are two main data compression classes: lossless and lossy. Compared to lossless compression, lossy compression can provide a much higher compression ratio by trading an acceptable accuracy loss. Lossy compressors for image and video have been well studied, such as JPEG~\cite{jpeg} and MPEG~\cite{le1991mpeg}, but they are human perception-driven rather than designed for scientific postanalysis and lack error-controlling mechanism.

Recently, a new generation of lossy compression for scientific data, especially floating-point data, has been developed, such as SZ~\cite{sz16, sz17, liang2018error}, ZFP~\cite{zfp}, MGARD~\cite{mgard}, and TTHRESH~\cite{ballester2019tthresh}.
Unlike lossy compressors for images and video, these lossy compressors provide strict error-controlling schemes, allowing users to control the accuracy loss in reconstructed data and even in post-analysis.

Considering the boom in GPU-based HPC systems and applications, SZ, ZFP, and MGARD are starting to roll out their GPU versions using CUDA~\cite{sanders2010cuda} (i.e., \cusz{}~\cite{cusz2020}, \cuzfp{}~\cite{cuZFP}, and \cudamgard{}~\cite{chen2021accelerating}), which provide much higher throughputs for compression compared with their CPU versions. \cuzfp{} (a transform-based compressor) allows the user to specify the desired bitrate (i.e., the average number of bits per value after compression), while \cusz{} (a prediction-based compressor) and \cudamgard{} (a multigrid-based compressor) allow the user to specify the maximum error that can be tolerated. \cuzfp{} with the fixed-rate mode can provide stably higher compression throughput, whereas \cusz{} and \cudamgard{} with the error-bounded mode tend to achieve a higher compression ratio.
In addition, there are also some optimization works based on these compressors to improve either compression ratio or compression throughput, which will be discussed in \SEC\ref{sec:related}.

\subsection{\cusz{} Framework}
\label{sec:cusz}

Since scientific data are mainly in floating-point representation, the randomly distributed bits in the exponent and mantissa are the major obstacle to significantly reducing the data size. This is because a change in floating-point values causes the exponent and mantissa representation to change from the most significant bit (MSB) to the least significant bit (LSB); even close values can have distinct bitsets. In comparison, a change in integer values results in fewer bit-level changes.

Thus, SZ framework converts the original floating-point data to integers in two stages: \Circled{1} it first \textit{predicts} the value of each data point using a prediction function such as Lorenzo predictor~\cite{ibarria2003out} and generates prediction errors (still floating-point values), which are the differences between the predicted and the original values, and \Circled{2} it then \textit{quantizes} the prediction errors to integers to reduce the bit randomness. After the prediction and quantization, lossless encoding works effectively on the integers (i.e., the approximation of prediction errors). The lossless encoding in SZ, such as gzip~\cite{gzip} or Zstd~\cite{zstd}, includes Huffman encoding and a dictionary encoding. In addition, {\cusz}, the GPU implementation of SZ framework, follows a similar compression pipeline with two primary adjustments in favor of performance, \Circled{1} it performs quantization on the original data before the prediction to remove the tight data dependency~\cite{cusz2020}, and \Circled{2} it omits the dictionary encoding stage.

\subsection{Dual-Quantization Method}\label{sub:bg-pq}

For the prediction and quantization stages, \cusz{} uses \emph{dual-quantization} method to achieve fine-grained parallelization; not only can chunked data blocks be compressed independently, but each data point can also be processed in parallel.
Specifically, \cusz{} first splits the whole dataset into multiple chunks. Then, it performs pre-quantization, Lorenzo prediction, and post-quantization. Note that pre-quantization is the only lossy stage (introducing compression errors) in the entire compression pipeline.

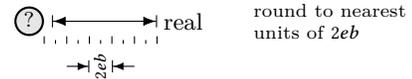
\begin{figure}[!htbp]
	\centering
	\begin{tikzpicture}[y=-1cm]

		\fontfamily{lmr}\selectfont

		\tikzset{symbolnode/.style={draw, circle, inner sep=-.5pt, minimum width=8pt
					, font=\strut
				}};

		\node[symbolnode, thick, fill=gray!20!white] (00)   at (-1.5, 1) {\small ?};
		\node[] (r) at (.55,1) {real};

		\draw[{|Latex}-{Latex}|] ([xshift=3pt]00.east) -- (.2, 1);

		\node[align=left, font=\footnotesize] at (2.5,1) {round to nearest \\units of $2eb$};

		\foreach \i in {0, 1, 2, 3, 4, 5 } {
				\draw (.2 - \i *0.3, 1.2) -- (.2 - \i *0.3, 1.3) ;
			}
		\foreach \i in { 1, 2, 3, 4, 5} {
				\draw (.2 + 0.15 - \i *0.3, 1.25) -- (.2 + 0.15 - \i *0.3, 1.3) ;
			}

		\draw [-{latex}|] (.2 - 1.2, 1.6) -- (.2  - 0.9, 1.6);
		\draw [|latex-] (.2 - 0.6, 1.6) -- (.2 - 0.3, 1.6);
		\node[rotate=90] at (-0.56, 1.6) {\scriptsize $2eb$};

	\end{tikzpicture}
	\caption{An illustration of error controlling in SZ.}
	\vspace{-4mm}
	\label{fig:2eb}
\end{figure}

We denote the input data as $d$ and the user-specified error bound as $eb$; the compression conducts the error-controlling process illustrated in \FIG~\ref{fig:2eb}. The error-boundness (i.e., the decompressed data error from the original is no greater than $eb$) can be guaranteed as
\begin{equation*}
	|\operatorname{round} ( d_i / (2\cdot eb) ) \times (2\cdot eb) - d_i | \le eb.
\end{equation*}
With the given parameter $r$, the output comprises two parts, quantization code $q' = q+r$ of limited numbers such that $q'= q-r$ and $-r < q < r$, and outlier that is out of range $(-r, r)$. We refer readers to the \cusz{} papers~\cite{cusz2020,cuszplus2021} for more details.

\subsection{Problem Statement}

While flourishing to achieve high data processing capabilities, GPU-based compressors target high versatility and a wide range of usage scenarios, such as in-memory compression~\cite{jin2021comet}, compression of MPI messages~\cite{zhou2021designing}, and reducing CPU-GPU data transfer time~\cite{besedin2015increasing}. On the one hand, the current cuZFP only allows it to use where high compression throughput is the priority, as its compression quality is low compared to \cusz{} under the same compression ratio. Moreover, \cuzfp{} does not support error-bounded mode.
\cudamgard{} can only provide very low compression throughput (will be shown in the evaluation).
On the other hand, \cusz's modularized design enables us to investigate/design new compression components and replace specific ones in the pipeline if needed. For example, the current \cusz{} is limited by its inefficient Huffman encoder\footnote{The Huffman encoding on the GPU includes building a large Huffman codebook and performing coarse-grained encoding based on the  Huffman tree.}.
As a result, in this work, we mainly focus on the error-bounded lossy compression framework \cusz{}, which can provide high compression ratios, and aim to drastically improve the compression throughput by designing a new high-performance GPU encoding approach to replace Huffman encoding in the pipeline.

In this work, we assume the data to compress is generated by scientific applications on GPUs, and then the compression would be directly performed on the data from the GPU memory; finally, the compressed data would be saved from GPUs to disks via CPUs.
There are several use cases that \thiswork{} targets. For example, it can reduce storage overhead as the compressed data will be saved from the GPU to the disk through the CPU for post-analysis.
It can also reduce memory overhead as the compressed data will be cached in the GPU global memory and decompressed on the GPU directly when the reconstructed data is needed for computation.

\section{Design of Proposed \thiswork}
\label{sec:design}

In this section, we present the design of our new GPU-based lossy compressor \thiswork{} with a series of optimizations.

\subsection{Overview of New Compression Pipeline}

Our proposed new compression pipeline is shown in \FIG~\ref{fig:system}. To fully utilize the computing power of GPUs, we aim to design a pipeline that maximizes parallelism while achieving high compression ratios by exploiting potential data patterns.

Inspired by \cusz{}, we adopt the dual-quantization method in the first stage of our compression pipeline for three reasons. \Circled{1} Its Lorenzo predictor exploits the spatial and dimensional information to reduce the entropy of input data significantly~\cite{sz17}. \Circled{2} It is fully parallelized, and its Lorenzo predictor is highly efficient due to the $O(n)$ time complexity. \Circled{3} Its quantization provides an error-controlling scheme for our lossy compression pipeline. However, the original quantization design in \cusz{} sets a threshold to separate regular quantization codes and outliers (\SEC\ref{sub:bg-pq}); though it favors a higher compression ratio, the amount of memory transaction hinders the performance, and hence it is not used in our design. Instead, we propose to optimize dual-quantization by neither shifting quantization codes nor handling outliers. Besides, we use the MSB to denote the sign of the data point. We will describe the detail of the optimized dual-quantization method in \SEC\ref{sub:prequantization}.

After that, we seek a new lossless encoding method that can provide high throughput and high compression ratios at the same time.
On the one hand, \cusz{} uses Huffman encoding that causes irregular memory access (i.e., the number of bits varies for each symbol). Thus, we look for a lossless encoding with more regular memory access. On the other hand, Huffman encoding achieves a high compression ratio but cannot handle sparse data (efficient prediction minimizes prediction errors in amplitude). Thus, it is critical to identify a representation for quantization codes that can expose as many continuous zero bits as possible.

\begin{figure}[t]
	\includegraphics[width=\linewidth]{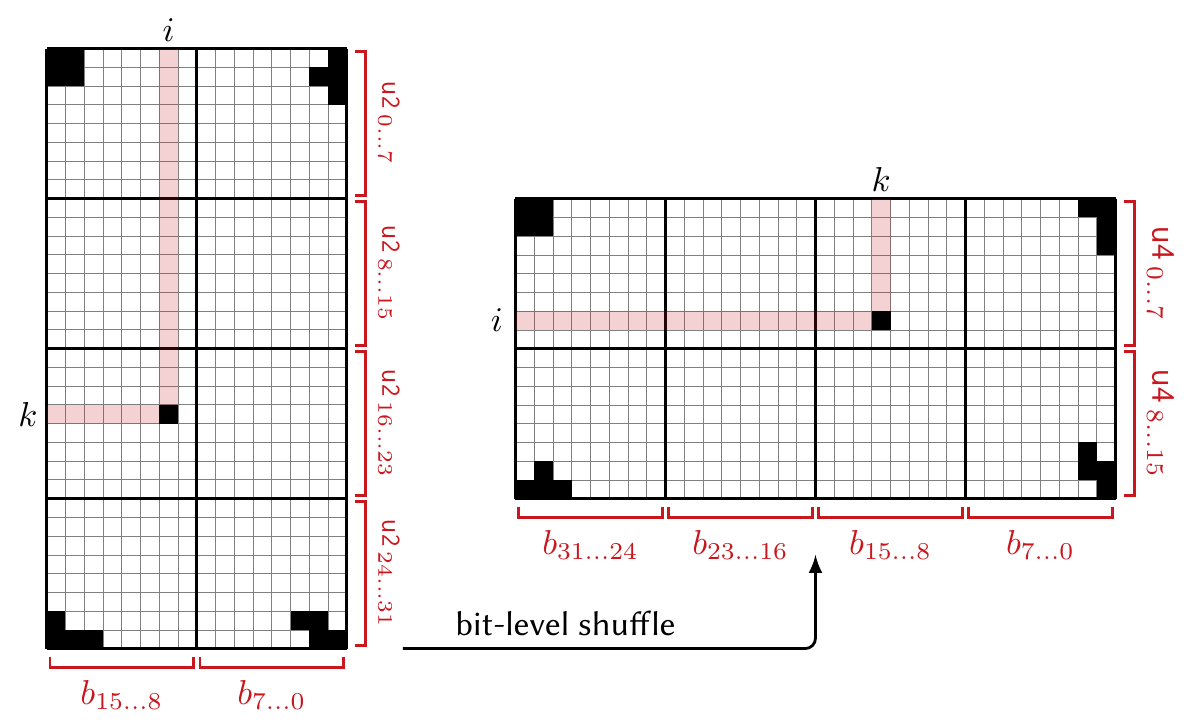}
	\caption{An illustration of bitshuffle algorithm.}
	\label{fig:general-Bitshuffle}
	\vspace{-2mm}
\end{figure}

To this end, we propose to adopt bitshuffle~\cite{masui2017bitshuffle} (as illustrated in \FIG~\ref{fig:general-Bitshuffle}) before performing encoding.
The advantages of using bitshuffle are twofold,
\Circled{1} it transforms the data representation to create more space redundancy for subsequent lossless compression, and
\Circled{2} it is a highly parallel process well-suited for GPU processing.
However, the bitshuffle operation is more time-consuming than the dual-quantization operation. Thus, we propose to optimize its performance by using warp-level functions and utilizing shared memory. We will present bitshuffle and our optimization in \SEC\ref{sub:Bitshuffle}.

Lastly, we propose a sparsification-style fast lossless encoding after bitshuffle. Specifically, we partition the data into many data blocks and then go through each data block to check if all values are zero: if so, we use a 0-bit to record the block; otherwise, we use a 1-bit to record it and copy this block to the output compressed array. However, this lossless encoding process is hard to achieve high performance on the GPU because the encoded address offsets are unknown for different data blocks. Therefore, we need to pre-compute the offset (i.e., the starting point of the memory address) and encode each data block according to its offset. The detail of our proposed lossless encoding method will be described in \SEC\ref{sub:encoder}.

Compared to \cusz{}, both \thiswork{} and \cusz{} use prediction and quantization to reduce the entropy of the datasets. However, the lossless encoding of our work is entirely different from \cusz{}. Instead of utilizing the time-consuming Huffman encoding to compress the quantization code (output of prediction and quantization), we propose to use a simple but effective pipeline with bitshuffle and our proposed lossless encoding. By doing this, the compression throughput is significantly increased, as shown in \FIG~\ref{fig:general-Bitshuffle}. Moreover, the compression ratio is no longer limited by Huffman encoding (an upper bound of 32), which means it is potentially increased. It is worth noting that we also modify the pre-quantization kernel to fit our pipeline by integrating the outliers and using the most significant bit to store the sign of the number, which also increases the throughput of the pre-quantization kernel.

\vspace{-1mm}
\subsection{Proposed Optimized Dual-Quantization}
\label{sub:prequantization}

We employ dual-quantization in the first stage of our compression pipeline because it can significantly reduce the entropy of input data by exploring the spatial correlation through the Lorenzo predictor~\cite{sz17}. Furthermore, its fine-grained parallelism with low time complexity (i.e., $O(n)$) further facilitates an efficient GPU implementation.
	{Moreover, its quantization provides an error-controlling scheme for lossy compression, which enables an error-bounded mode similar to \cusz{}. }
However, the original dual-quantization method handles outliers by compressing them separately and shifts all quantization codes by a radius for a higher compression ratio, which leads to throughput degradation. Therefore, we propose an optimized dual-quantization method for higher performance.
The main differences between the original and our optimized dual-quantization methods are threefold, \Circled{1} we remove shift operations to formulate values symmetrically distributed around zero, \Circled{2} we avoid separate handling of outliers for high performance, and \Circled{3} we use 1 bit to handle the sign of each quantization code instead of using 2's complement. We describe them in detail as follows.

First, we optimize the representation of quantization codes. According to our empirical analysis, although the original data type is a float of four bytes, most data can be denoted as less than four bytes after quantization. Thus, we propose to use two bytes to represent the quantization code, which indirectly achieves compression by transforming the data type.
Note that the out-of-range data points are very few compared to the whole dataset. Thus, even losing these elements' precision will not significantly affect the decompressed data quality, such as peak signal-to-noise ratio (PSNR).

Next, we optimize the mechanism that handles the outliers in the prediction.
The prediction in \cusz{} sets a threshold to distinguish outliers and normal data points. \cusz{} will compress outliers separately because the compression ratio of Huffman encoding in the next step depends on the entropy of the data. If the entropy is too high, Huffman encoding will need more bits to denote patterns. We note that it is unnecessary to separate the outliers and normal data points when replacing Huffman encoding with our proposed lossless encoder. Therefore, we propose to discard the outliers handling in our pipeline.

\begin{figure}[t]
	\centering
	\includegraphics[width=\linewidth]{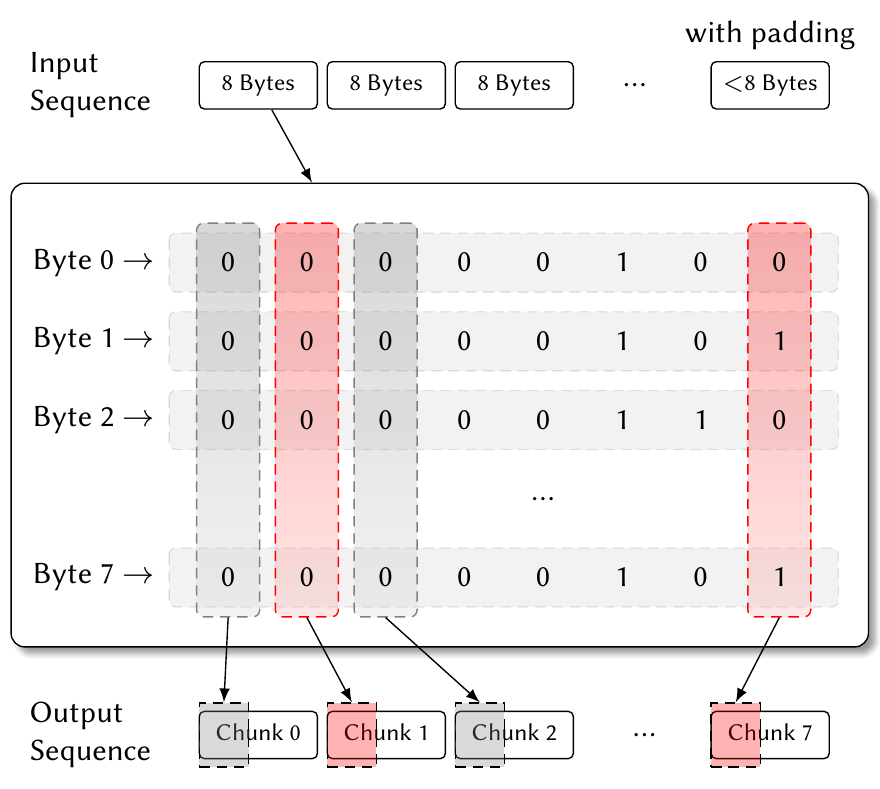}
	\caption{A simplistic fine-grained parallel bitshuffle~\cite{bitshuffle-gpu}.}
	\label{fig:Bitshuffle}
	\vspace{-2mm}
\end{figure}

Furthermore, we modify the negative numbers' data format to fit our encoding kernel's design. Specifically, instead of storing the data as a signed integer, we use an unsigned integer. We use the corresponding positive number with the most significant bit set as one for the negative number. This is because a negative number is represented as two's complement, consisting of many set bits when its absolute value is small.
This is unsuitable for our design because we expect the data bytes to have as many zeros as possible. To solve this issue, we propose to use the first bit of unsigned int to denote the positive and negative because the efficient prediction will keep the data in a small range around zero, which guarantees the valid number that two bytes can represent is more than enough for the quantization code. As aforementioned, few data points are out of range, so this modification accelerates the dual-quantization kernel due to fewer if-else branches and easier operations.

\subsection{Optimization of Bitshuffle on GPUs}
\label{sub:Bitshuffle}

Bitshuffle is an algorithm that re-organizes the dataset bit-wise by gathering the $n$-th bits of all the bytes in eight chunks. \FIG~\ref{fig:general-Bitshuffle} shows an example of bitshuffle. Bitshuffle fits our compression pipeline for two reasons: \Circled{1} it creates more spatial redundancy for the following lossless compression, and \Circled{2} there is no data dependency in bitshuffle, meaning it is highly parallelizable. However, the bitshuffle operation is time-consuming and needs bit-level operations. Therefore, we propose a series of optimizations to improve its performance, as described below.

The first optimization is fully leveraging shared memory in each thread block. Bitshuffle is a memory-intensive process that needs to access the same memory multiple times to get different bits of the same byte. Direct access to global memory has much higher latency than shared memory. Therefore, we propose to use shared memory to reduce the memory access overhead. Since we need to combine as many bits as possible to create more spatial redundancy at the bit level, we need to set shared memory size as large as possible to store more data. In our kernel design, we use a 32-by-32 array of unsigned integers, 4 bytes per array element (each element saves two quantization codes) to store the data. We set the thread block size to 32-by-32, corresponding to shared memory size. Note that the actual size of the 2D array in shared memory is 32-by-33 with padding to avoid bank conflicts.

After loading the data into shared memory, we need to extract the corresponding bits and put them together. This operation is challenging for GPU, since writing to the same memory location by all threads in a warp will cause data access conflicts. However, if we perform this 1-bit operation at a time, the parallelism advantage of GPU is much undermined; in other words, there would always be threads waiting for others to complete. To solve this issue, we use a warp-level vote function, \verb|__ballot_sync()| (requiring \verb|uint32_t| as the input type), to speed up this operation.
The vote function takes variable $v_a$ of each thread in a warp as input and outputs a \verb|uint32_t| $v_b$.
More specifically, $v_a$ of thread $i$ is used in the predicate to set $i$-th bit of $v_b$ with true (1) or false (0).
Therefore, we can extract certain bits of the element in the array and use the vote function to implement the shuffle process without sacrificing the parallelism.

\begin{figure}[t]
	\centering
	\includegraphics[width=\linewidth]{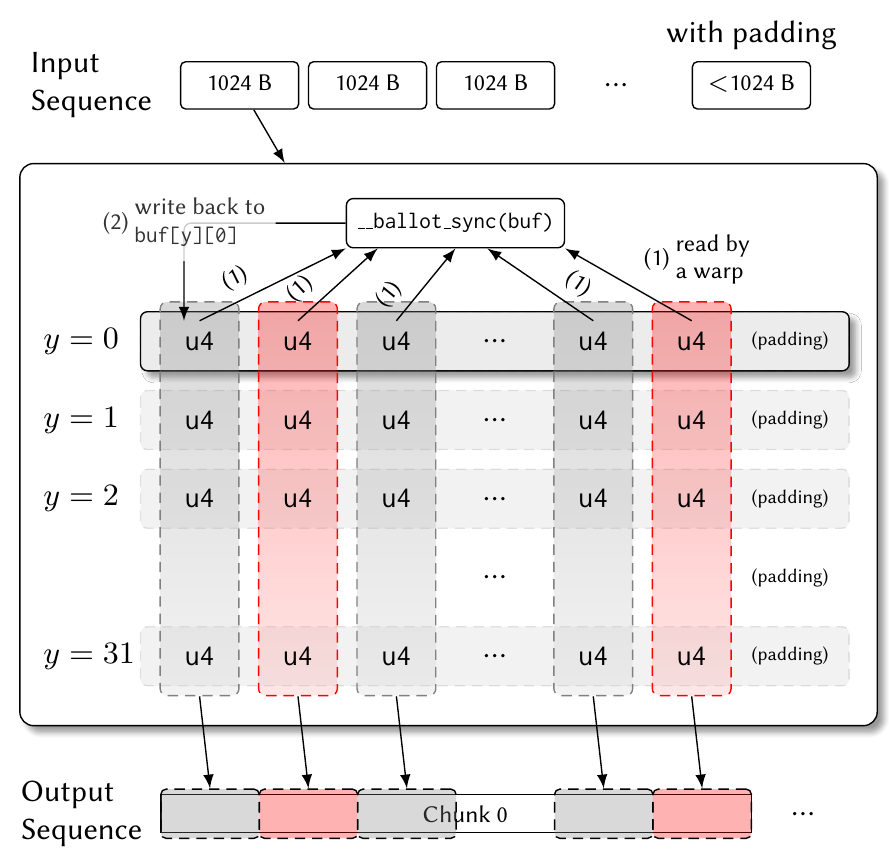}
	\caption{Our proposed scalable GPU bitshuffle method. \texttt{u4} stands for 4-byte unsigned integer type.}
	\label{fig:Bitshufflekernel}
\end{figure}

Then, we need to put the bitshuffled result back into the global memory to continue the encoding process. The simplistic way is to store the shuffled data independently in eight chunks, as shown in \FIG~\ref{fig:Bitshuffle}. However, the memory access of this simplistic solution is non-coalesced, which would cause a significant drop in throughput.
To solve this issue, we propose another optimization, as shown in \FIG~\ref{fig:Bitshufflekernel}. We store the result locally in the same thread block. The compression ratio will not be affected if the granularity is coarser than the following encoder. We process the data row-wise so that the bits in the same order are stored in the same column. We then write back column-wise. Note that padding allows us to access the data points column-wise without bank conflict.

\subsection{Proposed Fast GPU Lossless Encoder}
\label{sub:encoder}

\begin{figure}[t]
	\centering
	\includegraphics[width=\linewidth]{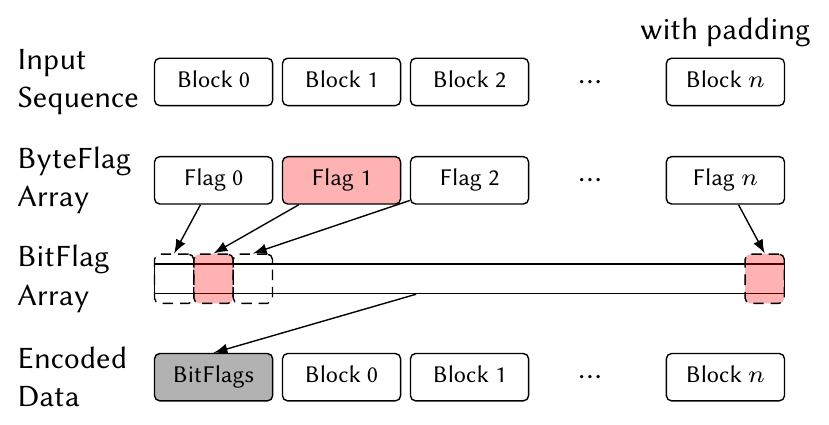}
	\caption{Our proposed fast GPU encoding method.}
	\label{fig:encodekernel}
\end{figure}

The prior study~\cite{masui2017bitshuffle} finds that bitshuffle works well with LZ4 lossless encoding on scientific floating-point data.
However, the LZ4 algorithm is unsuitable for GPU architectures~\footnote{LZ4 from nvCOMP~\cite{nvcomp-lz4} can only achieve 6.3~GB/s on our evaluation datasets.} due to the sequential nature of its search for repeated strings (similar to all LZ-family compression algorithms such as LZ77 and LZSS), leading to relatively low throughput.
To this end, we propose a new lossless encoding method to replace the LZ4 encoder and couple it with bitshuffle.
The overview of our proposed encoding method is presented in \FIG~\ref{fig:encodekernel}.
Specifically, this encoder has two phases. The first phase is to partition the data into data blocks and then iterate all data blocks to record whether all values in the data block are zeros or not in an array, called the ``bit-flag array''. Then, the second phase encodes the data based on the flag array generated in the first phase. During the encoding, if the corresponding bit flag is `0', we use this zero to denote it; if the bit flag is `1', we copy the whole block to our compressed data. This sparsification-style encoder is highly suitable for bitshuffle because bitshuffle reorganizes the representation of quantization codes at the bit level, creating many consecutive zero bits/bytes and hence zero blocks.

However, it is non-trivial to implement this encoding method efficiently on the GPU because the size of each data block after compression varies. Thus, we need to determine the memory offset
for every data block before encoding it.
We use the prefix-sum to compute the memory offsets.
To implement an efficient GPU prefix sum, we need device-wide synchronization to ensure sizes for all compressed blocks are ready. Two approaches can achieve this global synchronization. The first approach is to use the cooperative group API~\cite{cooperative-groups}. However, the maximum possible number of threads is limited and unsuitable for our problem. Another approach is to split one kernel into two since a synchronization can be conveniently triggered when a GPU kernel exits. Thus, we propose two phases in our encoding method, with the two corresponding optimized kernels detailed below.

To take advantage of the result stored in shared memory in the bitshuffle kernel to save the time to read again from global memory, we propose to fuse the bitshuffle and the first phase of our encoding in a single kernel. Note that while the granularity of the two processes is different, meaning that some threads in the kernel will be idle while others are executing, it is still more cost-effective than accessing global memory (which will be proved in the evaluation). Once the data is ready, we use statically allocated buffers in shared memory (\textcolor{R}{Lines 2--3}) to store the flags of each data block.
We then use fewer threads than in bitshuffle to iterate over the data blocks and record a flag indicating whether all data points in the same data block are zero (\textcolor{R}{Lines 14--16}).
The result will be temporarily stored in an \verb|uint8_t| array, called \verb|ByteFlagArr| (declared in \textcolor{R}{Line 3}).
Finally, we convert the byte-flag array to the bit-flag array through the bit-level operations (\textcolor{R}{Lines 18--20}).
Note that we also use the warp-level vote function to generate the bit-flag array to avoid data access conflicts (\textcolor{R}{Line 21}).
In addition, instead of wasting the byte-flag array, we will use it to calculate the prefix-sum for the offsets of data blocks in the second phase. The proposed fused kernel is shown below in detail.

\begin{lstlisting}[language=c++, keywords={__shared__, uint32_t, uint8_t, if, for}]
__shared uint32_t buf[32][33]
__shared uint32_t BitFlagArr[8]
__shared uint8_t  ByteFlagArr[256]
uint32_t cur
ltid = get_linear_threadid()

buf[Idx.y][Idx.x] = input[offset]; __syncthreads()

cur = buf[Idx.y][Idx.x]
for i in range(32):
  buf[Idx.y][Idx.x] = __ballot_sync(cur & (1U << i))
output[offset] = cur = buf[Idx.x][Idx.y]

if Idx.x * 4 < 32:
  for i in range(4):
    ByteFlagArr[ltid] = any(buf[Idx.x*4+i][Idx.y] != 0)
        
if Idx.y < 8:
  cur = ByteFlagArr[ltid]
  BitFlagArr[Idx.y] = __ballot_sync(buf)
   
WriteBackToGlobalMem(ByteFlagArr)
WriteBackToGlobalMem(BitFlagArr)
\end{lstlisting}

For the second phase, we directly call the high-performance ExclusiveSum function (\textcolor{R}{Line 1}) from \verb|NVIDIA::CUB| library~\cite{repo-nvidia-cub}. Then, we can obtain the memory offset of each compressed data block. After that, we launch our encode kernel to write the compressed data back to the output array in the global memory (\textcolor{R}{Lines 8-9}). Note that if the corresponding data block has a valid offset\footnote{The offset is valid if it is different from its previous offset.}, the compressed data block will be saved; otherwise, it will be discarded. The detail of our second kernel is shown below.

\begin{lstlisting}[language=c++, keywords={__shared__, uint32_t}]
PrefixSum(ByteFlagArr, PreSum)
__shared uint32_t sumArr[33]

ltid = get_linear_threadid()
SumArr[0] = PreSum[ltid]
SumArr[Idx.x+1] = PreSum[gid+1]

if SumArr[Idx.x+1] != SumArr[Idx.x]:
  output[offset] = input[ltid]
\end{lstlisting}

\section{Experimental Evaluation}
\label{sec:eval}

\subsection{Experimental Setup}
\label{sub:evalsetup}

\paragraph*{Platforms.}
We use two platforms in our evaluation:
\Circled{1} One node from an HPC cluster equipped with two 64-core AMD EPYC 7742 CPUs at 2.25GHz and four NVIDIA Ampere A100 GPUs (108 SMs, 40GB), running CentOS 7.4 and CUDA 11.4.120.
\Circled{2} An in-house workstation equipped with two 28-core Intel Xeon Gold 6238R CPUs at 2.20GHz and two NVIDIA GTX A4000 GPUs (40 SMs, 16 GB), running Ubuntu 20.04.5 and CUDA 11.7.99. {While we use one GPU for evaluation, multi-GPU processing is considered embarrassingly parallel with regard to single-GPU processing. This is because we partition data in a coarse-grained manner to fit into a single GPU, with a data chunk independent from another. With no data dependency, the multi-GPU comparison will only involve different numbers of data chunks.}

\begin{table}[ht]
	\sffamily
	\small
	\caption{\textcolor{B}{Real-world \texttt{float}-type datasets used in evaluation.}
	}
	\resizebox{.85\linewidth}{!}{
		\begin{tabular}{@{} >{\bfseries\scshape}lrr @{}}
			                                    & \bfseries\scshape\color{R} field data size & \bfseries\scshape\color{R} \#fields \\[-.4ex]
			\TABLETITLE datasets                &
			\TABLETITLE dimensions              &
			\TABLETITLE examples(s)                                                                                                \\
			\midrule
			\TABLECAPTION cosmology             & \TABLECAPTION 1,123.81 MB                  & \TABLECAPTION 6 in total            \\[-.3ex]
			HACC                                & 280,953,867                                & xx, vx                              \\
			\TABLECAPTION climate               & \TABLECAPTION 25.92 MB                     & \TABLECAPTION 70 in total           \\[-.3ex]
			CESM                                & 1,800$\times$3,600                          & CLDICE, RELHUM                      \\
			\TABLECAPTION cosmology             & \TABLECAPTION 536.87 MB                    & \TABLECAPTION 6 in total            \\[-.3ex]
			NYX                                 & 512$\times$512$\times$512                  & baryon\_density                     \\
			\TABLECAPTION climate               & \TABLECAPTION 100 MB                       & \TABLECAPTION 13 in total           \\[-.3ex]
			Hurricane                           & 100$\times$500$\times$500                  & CLDICE, QRAIN                       \\
			\TABLECAPTION quantum circuits      & \TABLECAPTION 630.74 MB                    & \TABLECAPTION 1 in total            \\[-.3ex]
			QMCPACK                             & 7,935$\times$69$\times$288                                  & einspline            \\
			\TABLECAPTION petroleum exploration & \TABLECAPTION 189.50 MB                    & \TABLECAPTION 16 in total           \\[-.3ex]
			RTM                                 & 449$\times$449$\times$235                  & snapshot\_1200                       \\
			\bottomrule
		\end{tabular}
	}
	\label{tab:datasets}
\end{table}

\paragraph*{Datasets.}
We conduct our evaluation and comparison based on six typical real-world HPC simulation datasets from the Scientific Data Reduction Benchmarks~\cite{zhao2020sdrbench}: HACC (cosmology particle simulation)~\cite{hacc}, CESM (climate simulation)~\cite{cesm-atm}, Hurricane (ISABEL weather simulation)~\cite{hurricane}, Nyx (cosmology simulation)~\cite{nyx}, QMCPACK (quantum Monte Carlo simulation)~\cite{QMCPACK}, and RTM (reverse time migration, seismic imaging for petroleum exploration)~\cite{jin2022improving}, which have been widely used in previous compression studies~\cite{wang2019compression, cusz2020, lu2018understanding, tian2021revisiting, cody2022optimizing, cuszplus2021, liu2021high,underwood2020fraz, barrow2022zhw,underwood2022optzconfig,liu2022dynamic}. {The details are shown in Table~\ref{tab:datasets}.}

Note that to compress particle datasets such as the HACC dataset with a minimum impact on the probability density function, prior work~\cite{di2018efficient} proposes to use point-wise relative error bound. To readily achieve that, an existing work~\cite{xincluster18} proposes to transform the original data using a logarithmic function and compress the log-transformed data with the corresponding absolute error bound (computed from the point-wise relative error bound). Thus, in this paper, we evaluate the log-transformed HACC dataset.

\vspace{-1mm}
\paragraph*{Baselines.}
We compare \thiswork{} with four state-of-the-art GPU lossy compressors, including \cuzfp{}~\cite{cuZFP}, \cusz{}~\cite{cusz2020}, \cudaSZx{}~\cite{szx}, and \cudamgard{}~\cite{chen2021accelerating}.
We exclude bitcomp~\cite{nvcomp-bitcomp} from the evaluation as it is closed-source software with an unknown compression algorithm.
	{We use five typical relative error bounds (relative to the value range of the data field), i.e., 1e$-$2, 5e$-$3, 1e$-$3, 5e$-$4, and 1e$-$4.}
Note that when comparing the compression throughput, we evaluate \cusz{}, \cudaSZx{}, and \cudamgard{} under the same error bound. In contrast, we evaluate \cuzfp{} under the same PSNR as ours as \cuzfp{} does not support the error-bounded mode.

\subsection{Evaluation Metrics}
\label{sub:metric-descri}
Our evaluation metrics include \Circled{1} compression ratio, \Circled{2} distortion between original and reconstructed data, \Circled{3} compression throughput, and \Circled{4} overall throughput, which are detailed as follows.
\begin{enumerate}[noitemsep, topsep=2pt, leftmargin=1.5em]
	\item \textit{Compression ratio} is one of the most commonly used metrics in compression research. It is a factor of the original data size to the compressed data size. Higher compression ratios mean denser information aggregation against the original data.
		      {It is worth noting that the bitrate is the average number of bits per value after compression. Since all of our evaluation datasets are single-precision floating-point data, the bitrate is calculated as 32 (bits) divided by the compression ratio.
			      We evaluate how each compressor corresponds to data quality at a specific bitrate, which will be presented in the rate-distortion curve in \SEC\ref{sub:eval-quality}.}
	\item \textit{Distortion} evaluation is crucial for evaluating lossy compression performance in data reconstruction quality. In this work, we mainly use PSNR to measure the distortion quality. Similar to prior work, we plot the rate-distortion curve for a fair comparison among different compressors and their diverse compression modes, which compares the distortion quality at the same bitrate.
		      {Moreover, we also adopt the Structural Similarity Index Measure (SSIM) to evaluate the reconstructed data quality. SSIM is a metric used to measure the similarity between two images. Details on the calculation can be found in~\cite{nilsson2020understanding}.}
	\item  \textit{Compression throughput} is how much data a compressor can process in one unit of time. It is a key advantage of using a GPU-based lossy compressor instead of a CPU-based one.
	\item \textit{Overall data-transfer throughput} is to measure the performance of transferring compressed data (through the network or CPU-GPU interconnect), including compression overhead. This metric is a composite indicator of compression ratio and speed. Higher compression ratio and higher compression throughput, higher overall data transfer throughput.
\end{enumerate}

\subsection{Evaluation of Compression Quality}
\label{sub:eval-quality}

\begin{figure}[t]
	\centering
	\includegraphics[width=\linewidth, trim={9pt 0 4pt 0}, clip]{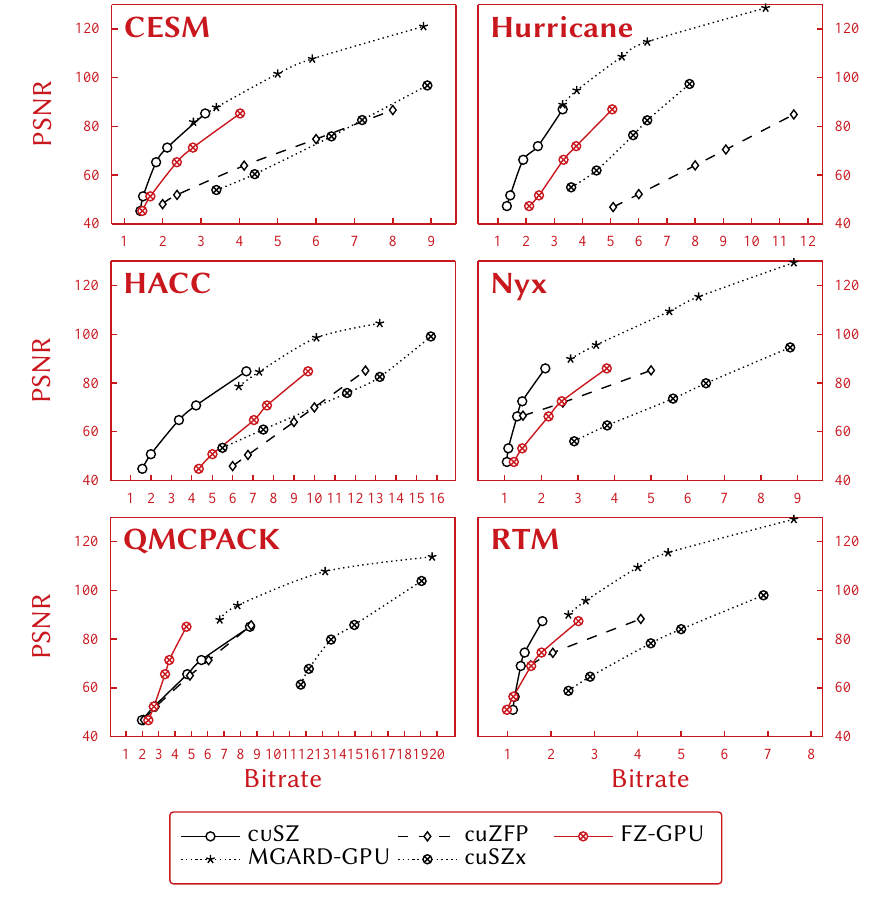}
	\caption{Rate-distortion of five GPU lossy compressors.}
	\label{fig:rate-distortion}
\end{figure}

\begin{figure*}[t]
	\centering
	\includegraphics[width=\linewidth]{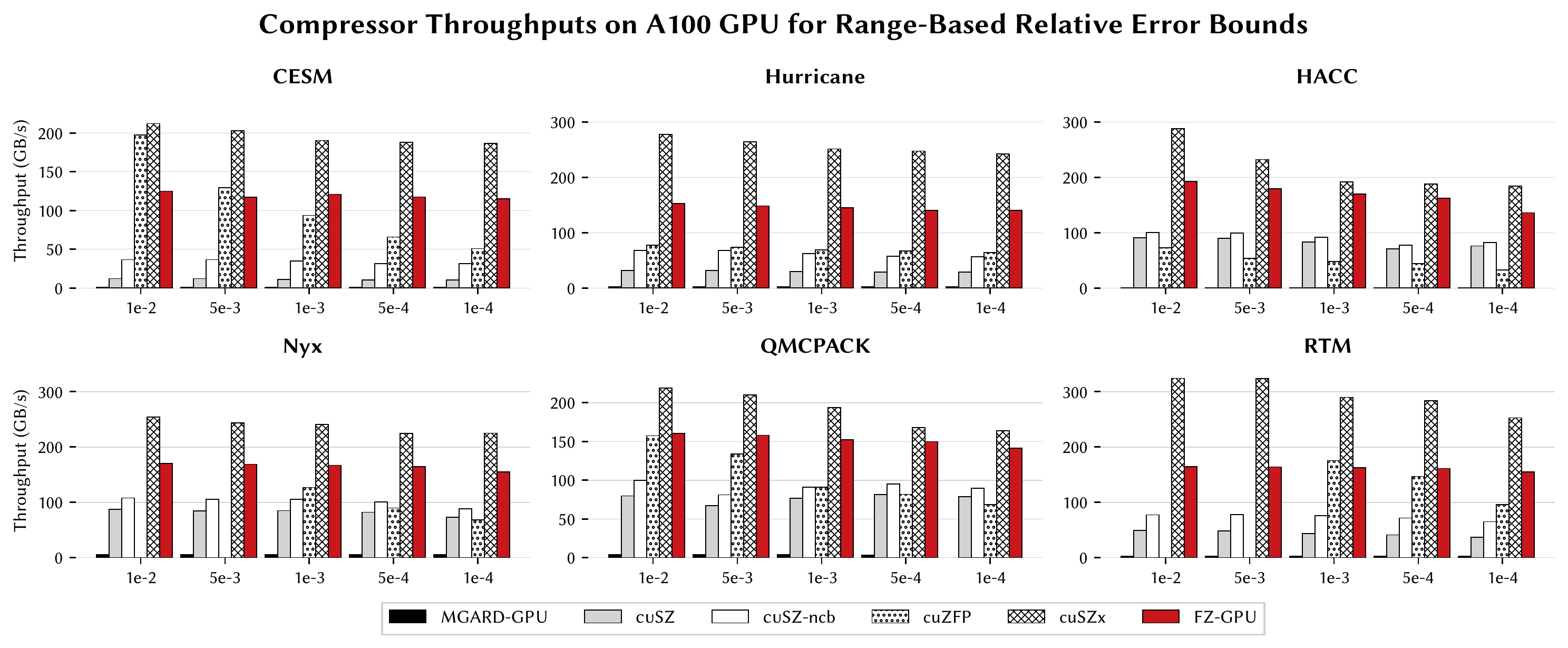}
	\caption{Compression throughput of \cuzfp, \cusz, \cusz-ncb (\cusz{} with no codebook building), \cudaSZx, \cudamgard, and \thiswork{} on NVIDIA Tesla A100. \cuzfp{}'s throughput corresponds to \thiswork{} with the same average PSNR.
	}
	\vspace{-2mm}
	\label{fig:a100throughput}
\end{figure*}

\begin{figure*}[t]
	\centering
	\includegraphics[width=\linewidth]{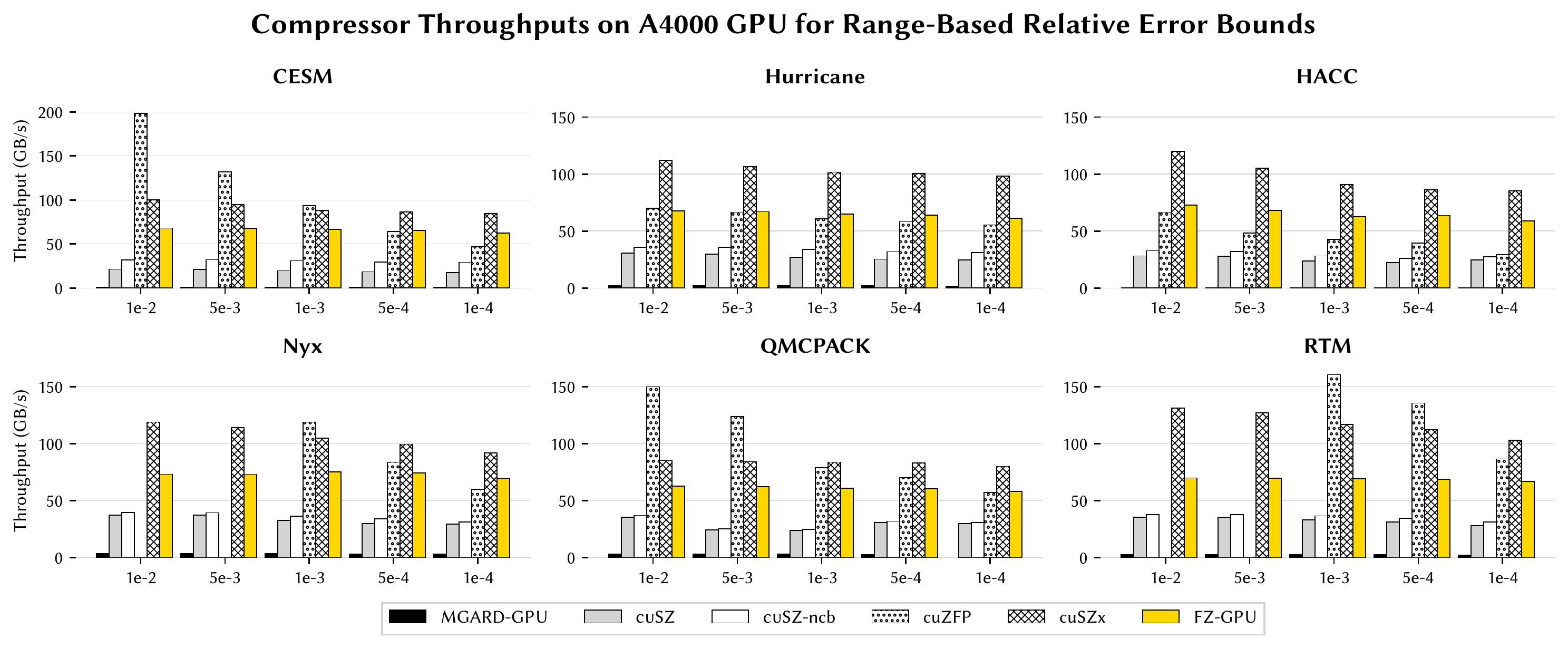}
	\caption{Compression throughput of \cuzfp, \cusz, \cusz-ncb (\cusz{} with no codebook building), \cudaSZx, \cudamgard, and \thiswork{} on NVIDIA RTX A4000. \cuzfp{}'s throughput corresponds to \thiswork{} with the same average PSNR.
	}
	\vspace{-2mm}
	\label{fig:a4000throughput}
\end{figure*}

First, we compare the five compressors' rate-distortion curves (i.e., distortion in PSNR versus bitrate), as shown in \FIG~\ref{fig:rate-distortion}.
Specifically, our platform uses different experimental settings to get the rate-distortion curves. We apply five different relative error bounds (relative to the value range) to \cusz{}, \cudamgard, \cudaSZx{}, and \thiswork. Due to the fact that \cuzfp{} does not support the error-bounded mode, we investigate a series of bitrates and select the bitrates with the same average PSNR as ours. Note that on Nyx and RTM, \cuzfp{} cannot achieve a similar PSNR as ours with the error bounds of 1e$-$2 and 5e$-$3. As shown in \FIG~\ref{fig:rate-distortion}, \thiswork{} has a similar compression ratio compared to \cusz{}. On the RTM dataset with high error bounds, the compression ratio of \thiswork{} is up to 1.1$\times$ higher than \cusz{} and 1.7$\times$ higher than \cuzfp{} on average.
Note that \thiswork{} has good stability regarding distortion. For example, on RTM with 400 timesteps, PSNR varies only from 86.1 dB to 87.5 dB under the relative error bound of 1e$-$4.
The analysis is detailed in the following sections.

\vspace{-1mm}
\paragraph*{Comparison with \cusz}

Since the lossy part (i.e., dual-quantization) of \thiswork{} is the same as \cusz, their PSNR is the same when we use the same error bound. Therefore, our bitrate is very close to \cusz{}. In some cases of the high error bound, \thiswork{} has a higher compression ratio than \cusz. For example, \FIG~\ref{fig:rate-distortion} shows that \thiswork{} has a $13.9\%$ improvement in compression ratio with an error bound of 1e$-$2 on the RTM dataset. This is because the RTM dataset contains many zero values and other highly smooth values. Therefore, after we apply bitshuffle, the shuffled data is mostly zero, resulting in a high compression ratio of our designed sparsification-like encoding method.
In comparison, \cusz{} has the compression ratio upper bounded by 32 due to Huffman encoding. Moreover, \cusz{} does not fully utilize the spatial redundancy of the RTM dataset. In contrast, our lossless encoder guarantees that the spatial redundancy is effectively compressed and the compression ratio is up to 128. Thus, the smooth values on the RTM dataset make the bitshuffled data more suitable for our lossless encoder.
We note that \cusz{} cannot work correctly on 3D QMCPACK due to a Huffman encoding error; therefore, we apply \cusz{} on the 1D QMCPACK (flattened) for a comparison.

\vspace{-1mm}
\paragraph*{Comparison with cuZFP}
\thiswork{} achieves a much higher compression ratio under the same average PSNR on almost all datasets compared to \cuzfp, except for some high error-bound cases on Nyx and RTM. For example, \cuzfp{} has a compression ratio of 21.3 on Nyx with the error bound of 1e$-$2, while \thiswork{} has a compression ratio of 14.5. This is because the two datasets under high error bounds are very smooth (most quantization codes are zeros), where \cuzfp{} is highly effective. But \cuzfp{} loses this advantage quickly when the error bound is lower. That is because the lower error bound gives the Nyx and RTM datasets higher entropy (like other datasets) after dual-quantization. The compression method of \cuzfp{} cannot handle such a complex dataset effectively.

\vspace{-1mm}
\paragraph*{Comparison with \cudamgard}
We note that due to memory issues, \cudamgard{} cannot work correctly on 1D datasets. For example, it cannot compress HACC on A100 with the relative error bound of 1e$-$4 and on A4000 with all the relative error bounds. Also, \cudamgard{} fails to compress QMCPACK with the error bound of 1e$-$4 because the compressed size is larger than the original size. As a result, the rate-distortion curve in \FIG~\ref{fig:rate-distortion} only contains 4 points for \cudamgard{} on QMCPACK and HACC.
\FIG~\ref{fig:rate-distortion} shows that under the same relative error bound, \cudamgard{} has higher PSNR on all datasets because
\cudamgard{} over-preserves the data distortion.
Regarding rate-distortion, \cudamgard{} is similar to \cusz{} and slightly better than \thiswork{} on CESM, Hurricane, and Nyx, since it uses a multi-grid-based approach with high time complexity (a large coefficient before $O(N)$) to achieve an accurate approximation.
\thiswork{} is close to \cudamgard{} on RTM. For example, \thiswork{} has a bitrate of 2.6 at the error bound of 1e$-$4, while \cudamgard{} has a bitrate of 2.4 at the error bound of 1e$-$2, with similar PSNR values. On QMCPACK, \thiswork{} outperforms \cudamgard{}: specifically, \thiswork{} has a bitrate of 4.7 at the error bound of 1e$-$4, while \cudamgard{} has a bitrate of 6.7 at the error bound of 1e$-$2, again with similar PSNR values.

\vspace{-1mm}
\paragraph*{Comparison with \cudaSZx}
Under the same relative error bound, \thiswork{} has a much higher compression ratio than \cudaSZx. Specifically, \thiswork{} has an average compression ratio improvement of 2.4$\times$ and 4.3$\times$ higher compression ratio than \cudaSZx{} at most on the QMCPACK dataset with a relative error bound of 1e$-$2. Although \cudaSZx{} has higher PSNR than \thiswork{} under the same error bound, \thiswork{} has a higher compression ratio under similar PSNR according to the curve shown in \FIG~\ref{fig:rate-distortion}.
This is because \thiswork{} employs the Lorenzo predictor (to reduce the entropy of the input data) and minimizes the \textit{bitwise} data redundancy, whereas \cudaSZx{} only reduces the \textit{blockwise} data redundancy.

\subsection{Evaluation of Compression Throughput}
\label{sub:eval-throughput}

Next, we evaluate the compression throughput of these three methods on A100 and A4000 GPUs, as shown in \FIG~\ref{fig:a100throughput} and \FIG~\ref{fig:a4000throughput}, respectively. Specifically, we measure their kernel time and apply five relative error bounds to \cusz{} and \thiswork{}.
Due to the fact that \cuzfp{} does not support the error-bounded mode, we investigate a series of bitrates and select the bitrates that have the same average PSNRs as ours.
Then, we use these bitrates to get the corresponding compression throughput. {In some cases, since cuZFP cannot achieve similar PSNR, we use fewer bars to display valid results.}
\FIG~\ref{fig:a100throughput} illustrates that on A100, \thiswork{} achieves a speedup of up to 11.2$\times$ over \cusz, and a speedup of up to 4.2$\times$ over \cuzfp.
On the A4000 platform, \FIG~\ref{fig:a4000throughput} shows that \thiswork{} achieves a speedup of up to 3.6$\times$ over \cusz{}, and a speedup of up to 2.0$\times$ over \cuzfp{}. Note that \cusz{} includes all \cusz's kernels, while \cusz-ncb does not include the time to generate the Huffman codebook (as this part can be done on the CPU). {Note that the decompression pipeline is highly symmetrical to the compression pipeline, exhibiting throughput nearly identical to that of compression. Due to space constraints, we do not present a detailed evaluation.}

\vspace{-1mm}
\paragraph*{Comparison with \cusz{}}
On A100, we observe that \thiswork{} has higher throughput than \cusz{} on all datasets. \thiswork{} has an average speedup of 4.2$\times$ than \cusz{} overall. on the CESM dataset, \thiswork{} achieves an even higher average speedup of 10.7$\times$. The performance compared with \cusz{} on CESM is better than other datasets because the Huffman codebook generating time in \cusz{} is almost the same among all datasets. For datasets like CESM, the field size is smaller than others, and the throughput of Huffman codebook generation is relatively lower. The result of \cusz-ncb also proves this; the ratio of \cusz-ncb to \thiswork{} is around 0.5 on almost all datasets. In contrast, \thiswork{} is highly stable across different datasets because our bitshuffle and fast encode kernels have almost the same amount of operations for the same data size.

The result on A4000 also demonstrates that \thiswork{} has more stability and higher throughput on different datasets compared to \cusz. Our throughputs are consistently around 70~GB/s, an average of 2.4$\times$ higher than \cusz{}. However, it is unusual that \cusz{} on A4000 has a higher throughput for the CESM dataset than on A100. This is due to the CESM dataset's small data size per field (i.e., 24.7 MB), which is sufficient for A4000 to warm up but not for A100. This phenomenon is less pronounced with datasets having larger field sizes than CESM.

\begin{figure*}
	\centering
	\includegraphics[width=.95\linewidth]{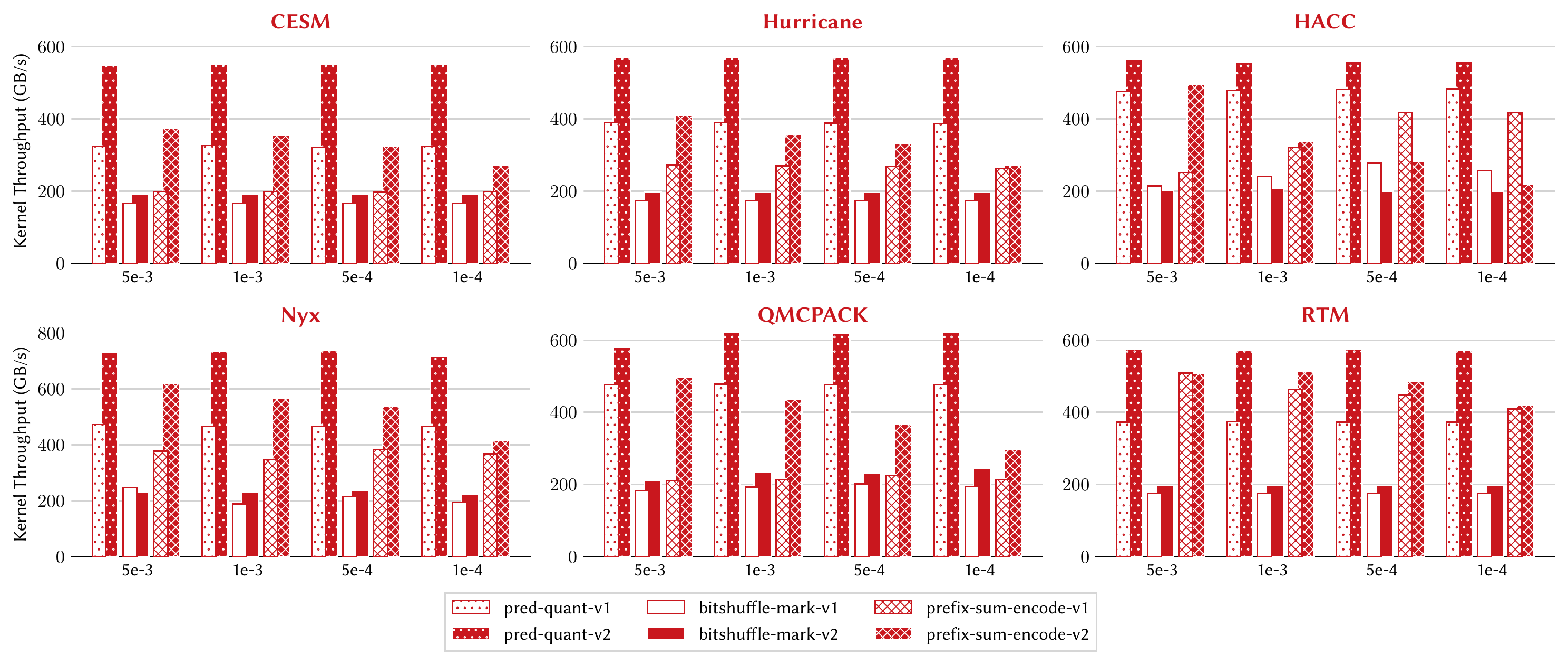}
	\caption{Performance improvements of our proposed optimizations for different compression kernels on NVIDIA A100.}
	\vspace{-2mm}
	\label{fig:optmize-kernels}
\end{figure*}

\vspace{-1mm}
\paragraph*{Comparison with cuZFP}
On A100, \thiswork{} has an average speed-up of 2.3$\times$ than \cuzfp{}. Furthermore, we observe that \thiswork{} achieves higher throughput on almost every experimental setting, except for the high error-bound cases on CESM and RTM. For example, on CESM with the error bound of 1e$-$2, our throughput is 125.0~GB/s, while the throughput of \cuzfp{} is 197.6~GB/s. The reason is that \cuzfp{} employs discrete cosine transform and bit truncation, which can be efficiently performed on the GPU by matrix operations, and achieves high efficiency when the data is super smooth (as aforementioned, RTM and CESM after pre-quantization have many zero values with a high error bound). But this advantage of \cuzfp{} disappears quickly as the error bound becomes lower.

On A4000, \thiswork{} has an average speedup of 1.3$\times$ than \cuzfp{}. However, we notice that the throughput of \cuzfp{} maintains almost the same between A4000 and A100. This is because \cuzfp{} is limited by GPU memory bandwidth rather than peak performance.

\vspace{-1mm}
\paragraph*{Comparison with \cudamgard}
As mentioned in \SEC\ref{sub:eval-quality}, MGARD-GPU cannot work correctly on 1D datasets due to memory issues. Moreover, although it can work on A100 in some error bounds, the compression throughput is unsteadily low (e.g., 0.018~GB/s on 1D HACC with 1e$-$2 error bound). With excluding the extreme cases, \thiswork{} averages 87.0$\times$ and 45.7$\times$ in throughputs, compared to \cudamgard. We also note that \cudamgard{} does not scale well from A4000 to A100. For example, its throughput on CESM with the relative error bound of 1e$-$2 is 0.62~GB/s on A100 and 0.67~GB/s on A4000, far less distinguishable than the hardware specifications. This demonstrates that \cudamgard{} does not respond well to different grades of modern GPUs.

\begin{figure*}[ht]
	\centering
	\includegraphics[width=.95\linewidth]{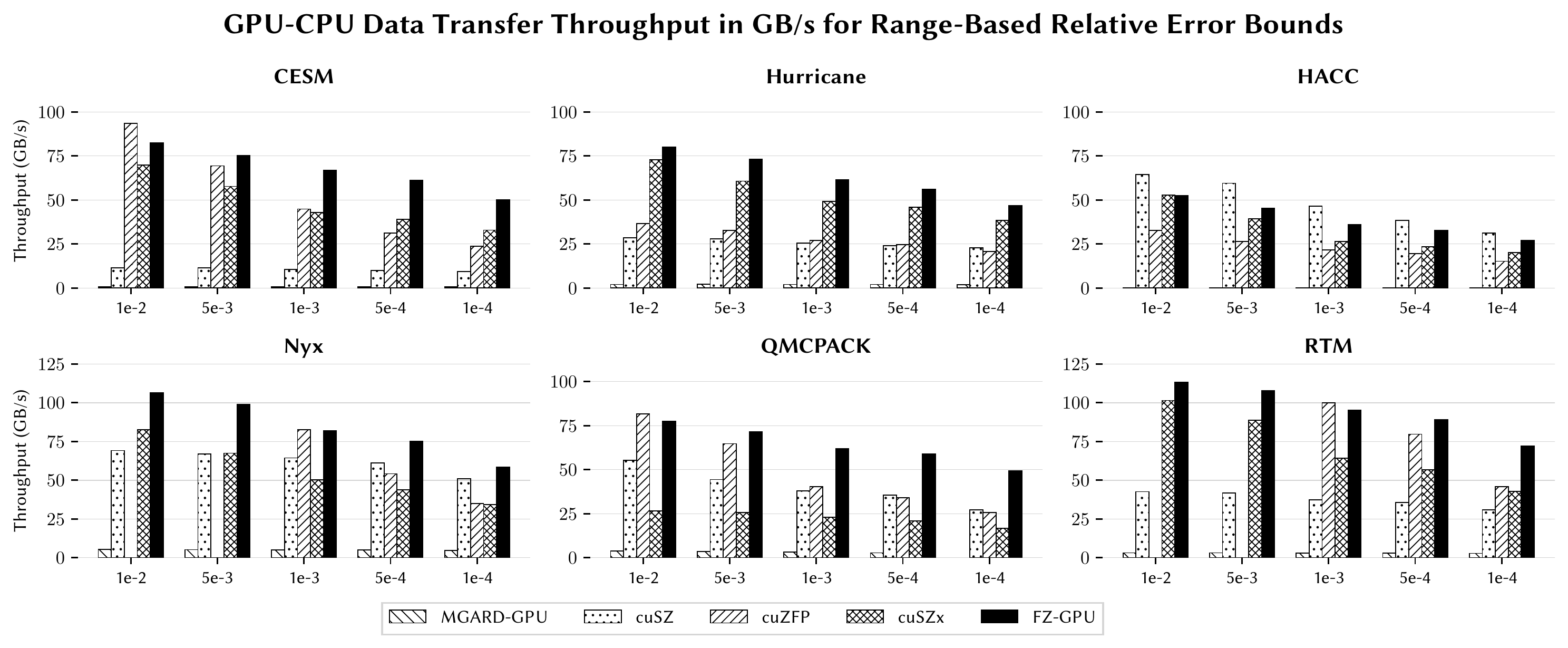}
	\caption{Overall CPU-GPU data-transfer throughput of \cuzfp, \cusz, \cudaSZx, \cudamgard, and \thiswork{} on NVIDIA A100.
	}
	\vspace{-2mm}
	\label{fig:overallthroughput}
\end{figure*}

\paragraph*{Comparison with \cudaSZx}
On A100, \cudaSZx{} has higher compression throughput on all datasets.
	{The compression throughput of \cudaSZx{} is 1.5$\times$ higher than \thiswork{} in average. This is because \cudaSZx{} uses a straightforward compression pipeline,
		which divides the input data into blocks and handles the constant blocks and non-constant blocks separately. This makes \cudaSZx{} highly efficient but also results in a fairly low compression ratio (as illustrated in \SEC\ref{sub:eval-quality}).} Note that the throughput of \cudaSZx{} on QMCPACK is relatively lower compared with other datasets. This is because the QMCPACK dataset consists of many unsmooth floating data points. Thus, non-constant blocks are much more than constant blocks.
In contrast, \thiswork{} is highly stable over different datasets.
On A4000, \cudaSZx{} also has advantages in throughput; the improvement is the same as that on A100, which is 1.5$\times$ on average.

\vspace{-1mm}
\paragraph*{Comparison with the CPU implementation.} We also implement our proposed lossy compression algorithm on multi-core CPUs using OpenMP (called ``FZ-OMP'')  and compare it with \thiswork{}.
The evaluation results show that \thiswork{} with A100 has speedups of 38.8$\times$, 42.4$\times$, 36.3$\times$, 31.8$\times$, 34.8$\times$, and 37.6$\times$ over FZ-OMP with Intel Xeon Gold 6238R CPUs (32 cores/threads) on HACC, CESM, Nyx, Hurricane, QMCPACK, and RTM, respectively.
Moreover, FZ-OMP has higher throughput than SZ-OMP due to its efficient compression algorithm. For example, the average throughput of FZ-OMP is 1.7$\times$, 2.5$\times$, and 2.0$\times$ higher than that of the original SZ-OMP (v.2.1.12.5)~\cite{repo-sz-omp} on the 3D Hurricane, Nyx, and RTM datasets, respectively, with the Intel CPUs using 32 cores/threads~\footnote{Note that the performance of both FZ-OMP and SZ-OMP increases as the number of threads increases to 32 (with up to 21.2$\times$ speedup), but it does not increase much with more than 32 threads on some datasets due to the limited workload per core.} (SZ-OMP only supports 3D data).
This demonstrates that both our proposed GPU performance optimizations and our compression algorithm contribute to the significant performance improvement over SZ/\cusz.

\subsection{Evaluation of Proposed Optimizations}
Finally, we present the evaluation of each of our optimizations in detail. The breakdown of the performance improvement on A100 is illustrated in \FIG~\ref{fig:optmize-kernels}.
Different bars represent different versions of each compression kernel, detailed as follows:
\begin{enumerate}[noitemsep, topsep=2pt, leftmargin=1.3em]
	\item \textbf{pred-quant-v1}: The original dual-quantization kernel.
	\item \textbf{pred-quant-v2}: Our optimized dual-quantization kernel without shifting and outlier handling.
	\item \textbf{bitshuffle-mark-v1}: Two separate kernels for bitshfuffle and mark operations.
	\item \textbf{bitshuffle-mark-v2}: One fused kernel for both bitshfuffle and mark operations.
	\item \textbf{prefix-sum-encode-v1}: Our prefix-sum \& fast encode kernel.
	\item \textbf{prefix-sum-encode-v2}: The same kernel as v1, while the encoding is improved due to dual-quantization optimization.
\end{enumerate}

\FIG~\ref{fig:optmize-kernels} shows that the dual-quantization kernel has a speedup of up to 1.7$\times$ because we remove the branches in the original kernel. More specifically, GPU executes instructions at the warp level, and different branches incur warp divergence, which is resolved sequentially. The kernel fusion of bitshuffle and bit-flag array also brings a speedup of up to 1.1$\times$. The kernel fusion can avoid extra access to the global memory by directly caching the bitshuffled data in shared memory. Our prefix-sum-encode kernel also has a speedup of up to 1.9$\times$ because of dual-quantization optimization. This is because fewer data blocks are encoded, so the encoding time is much lower than before. In addition, on the HACC dataset, the encoding time differs from other datasets (i.e., v1 has higher throughput than v2). This is because Lorenzo prediction is less effective for unsmoothed data like HACC; it generates many large irregular integers, affecting the encoding performance.
In contrast, \cusz{} does not have this phenomenon as it otherwise handles these irregular integers as outliers.

\begin{figure*}[t]
	\centering
	\setlength{\fboxsep}{0pt}
	\includegraphics[width=.93\linewidth,
		trim={0 5pt 0 3pt},
		clip=true
	]{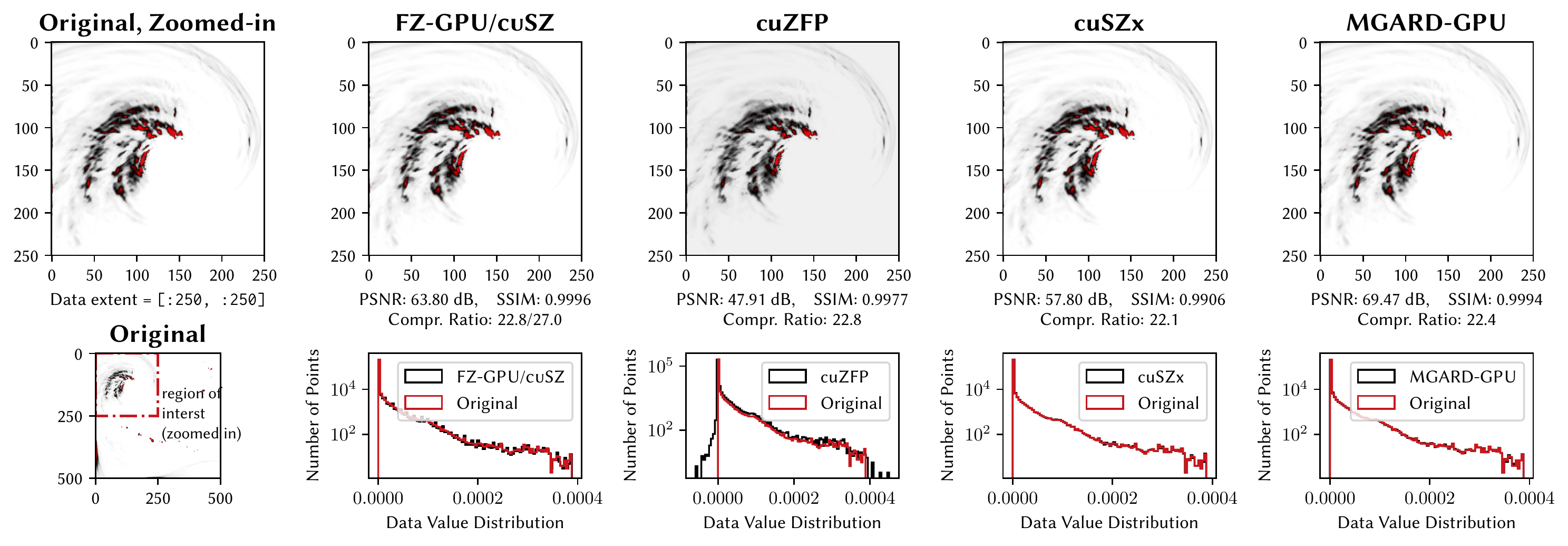}
	\caption{Reconstructed data quality using various GPU-based lossy compressors on field \texttt{QSNOWf48} (slice 50) in the Hurricane dataset, under a similar compression ratio. The first row shows the visualization of the region of interest, while the second row shows the data distribution comparison between the decompressed and the original data for each compressor.}
	\vspace{-4mm}
	\label{fig:ssim}
\end{figure*}

\subsection{Evaluation of Overall Throughput}
Besides compression throughput ($T_{\text{compr}}$), overall throughput considering the time in moving compressed data between GPU and CPU, is also a critical metric for overall application performance.
Thus, we propose to use this metric to further evaluate the efficiency of different compressors in practice. Specifically, the overall throughput can be calculated as
$$T_{\text{overall}}= \big( (BW\times \text{CR})^{-1} + T_{\text{compr}}^{-1}  \big)^{-1},$$
where $BW$ is the memory bandwidth between GPU and CPU and $CR$ is the compression ratio.
Our HPC cluster node is equipped with 4 A100 GPUs connected to the CPU via a 32-lane PCIe 4.0 interconnect; each GPU can leverage up to 16-lane bandwidth (i.e., 32~GB/s).
Based on our benchmarking result using~\cite{repo-gpu-cpu-benchmarking}, when the 4 GPUs read/write data from/to the CPU simultaneously, the bandwidth for each GPU can be as low as 11.4~GB/s (aggregately about 45~GB/s). Finally, we measure the overall data-transfer throughputs of different compressors and show them in \FIG~\ref{fig:overallthroughput}.
It illustrates that \thiswork{} achieves the best overall throughput on almost all datasets at all evaluated relative error bounds. Note that for the interconnections with effective bandwidth lower than 15~GB/s (e.g., networks), \thiswork{} method can achieve the optimal balance of compression ratio and throughput.
We leave the evaluation in node communication for future work.

\subsection{Evaluation of Reconstructed Data Quality}
Finally, we use \FIG~\ref{fig:ssim} to demonstrate the reconstructed data quality for all five lossy compressors by utilizing PSNR and SSIM. We select a similar compression ratio at approximately $22.8\times$ for a fair comparison, with different error bounds or bitrate configured.

Specifically, \thiswork{} has the identical reconstructed data quality to that of \cusz{} because of the shared error control scheme in our pipeline; thus, they share the same data visualization. Moreover, \thiswork{} has the highest SSIM among all compressors. Given that SSIM is a metric designed based on image structure, contrast, and luminance~\cite{setiadi2021psnr}, this demonstrates that \thiswork{} has a higher capability to preserve the quantity of interests or features than other compressors. On the other hand, the PSNR of our proposed pipeline is much higher compared to \cuzfp{} and \cudaSZx{} under a similar compression ratio. Although the PSNR of \thiswork{} is slightly lower than \cudamgard{}, \cudamgard{} has a very low throughput (4.9~GB/s compared to 65.4~GB/s in \thiswork{}). This is because \cudamgard{} uses a multi-grid-based approach with high time complexity (a large coefficient before $\mathcal{O}(N)$) to achieve an accurate approximation.

\section{Related Work}
\label{sec:related}

Some GPU-based lossy compression works have been optimized for scientific data, with a focus on CUDA architectures~\cite{sanders2010cuda}. For example,
\cusz{} is the first lossy compression framework that provides the error-bounded mode (detailed in \SEC\ref{sec:cusz}).
\cuzfp{} is the CUDA implementation of ZFP algorithm~\cite{zfp}, which performs near orthogonal transform and bit truncation over the split blocks of the data.
Tian et al.~\cite{cuszplus2021} proposed using run-length encoding in place of Huffman encoding to improve the compression ratio of \cusz{} for high error-bound scenarios. Yu {et al.} proposed \cudaSZx{}~\cite{szx} based on the \cusz{} framework that achieves very high compression throughput by using lightweight bitwise operations. Chen et al. developed \cudamgard{}~\cite{chen2021accelerating} that optimizes data refactoring kernels for GPU accelerators to enable efficient creation and manipulation of data in multigrid-based hierarchical forms.
Bitcomp~\cite{nvcomp-bitcomp} is a proprietary lossy compression developed by NVIDIA for scientific data, which has a similar performance as \cudaSZx{}.

In addition to lossy compression on GPUs, there are some GPU-based lossless compression works for scientific data. For example, Tian et al.~\cite{tian2021revisiting} proposed and implemented an efficient Huffman encoding approach for modern GPU architectures to parallelize Huffman encoding algorithm and utilize the GPU's high memory bandwidth.
Rivera et al.~\cite{cody2022optimizing} bitwise a deep architectural optimization for two Huffman decoding algorithms to take advantage of CUDA GPU architectures.
Knorr et al.~\cite{knorr2021ndzip} proposed an efficient GPU lossless compression of scientific floating-point data on GPUs using integer Lorenzo transform and vertical bit packing.
Masui et al.~\cite{masui2017bitshuffle} proposed a CPU vectorized compression using bitshuffle and LZ4, and~\cite{bitshuffle-gpu} is its simple GPU implementation.

\section{Conclusion and Future Work}
\label{sec:conclusion}
In this paper, we develop a fast and high-ratio error-bounded lossy compressor on GPUs for scientific data.
Specifically, we design a new compression pipeline that consists of dual-quantization, bitshuffle, and fast lossless encoding. We also propose a series of architectural optimizations for each GPU compression kernel, including warp-level optimization for bitwise operations, maximization of shared memory utilization, and multi-kernel fusion. Finally, we evaluate our proposed \thiswork{} on six representative scientific datasets and demonstrate its high compression throughput and ratio.

In the future, we plan to \Circled{1} exploit fusing all GPU kernels into one to improve the performance further,
\Circled{2} adapt \thiswork{} to other GPU platforms by using code translation tools such as HIPFY~\cite{repo-hipfy} for AMD GPUs and SYCLomatic~\cite{repo-syclomatic} for Intel GPUs,
and \Circled{3} evaluate \thiswork{} with real-world applications requiring fast compression, such as memory compression.

\section*{Acknowledgment}
This research was supported by the Exascale Computing Project (ECP), Project Number: 17-SC-20-SC, a collaborative effort of two DOE organizations---the Office of Science and the National Nuclear Security Administration, responsible for the planning and preparation of a capable exascale ecosystem, including software, applications, hardware, advanced system engineering and early testbed platforms, to support the nation's exascale computing imperative. The material was supported by the U.S. Department of Energy, Office of Science, Advanced Scientific Computing Research (ASCR), under contract DE-AC02-06CH11357.
This work was also supported by the National Science Foundation under Grants OAC-2003709, OAC-2104023, OAC-2303064, OAC-2247080, and OAC-2312673.
This research was also supported in part by Lilly Endowment, Inc., through its support for the Indiana University Pervasive Technology Institute.

\clearpage
\renewcommand*{\bibfont}{\footnotesize}
\printbibliography[]

\end{document}